\documentclass[12pt,letterpaper]{article}
\usepackage{mystyle}
\begin{document}

\title{Secure Degrees of Freedom of $K$-User Gaussian Interference Channels: A Unified View\thanks{This work was supported by NSF Grants CNS 09-64632, CCF 09-64645, CCF 10-18185 and CNS 11-47811.}}

\author{Jianwei Xie \qquad Sennur Ulukus\\
\normalsize Department of Electrical and Computer Engineering\\
\normalsize University of Maryland, College Park, MD 20742 \\
\normalsize {\it xiejw@umd.edu} \qquad {\it ulukus@umd.edu}}

\maketitle

\begin{abstract}

We determine the exact sum secure degrees of freedom (d.o.f.) of the
$K$-user Gaussian interference channel. We consider three different
secrecy constraints: 1) $K$-user interference channel with one
external eavesdropper (IC-EE), 2) $K$-user interference channel with
confidential messages (IC-CM), and 3) $K$-user interference channel
with confidential messages and one external eavesdropper (IC-CM-EE).
We show that for all of these three cases, the exact sum secure
d.o.f.~is $\frac{K(K-1)}{2K-1}$. We show converses for IC-EE and
IC-CM, which imply a converse for IC-CM-EE. We show achievability for
IC-CM-EE, which implies achievability for IC-EE and IC-CM. We develop
the converses by relating the channel inputs of interfering users to
the reliable rates of the interfered users, and by quantifying the secrecy penalty in terms of
the eavesdroppers' observations. Our achievability uses structured
signaling, structured cooperative jamming, channel prefixing, and
asymptotic real interference alignment. While the traditional
interference alignment provides some amount of secrecy by mixing unintended
signals in a smaller sub-space at every receiver, in order to attain the optimum sum
secure d.o.f., we incorporate structured cooperative jamming into the
achievable scheme, and intricately design the structure of all of the
transmitted signals jointly.
\end{abstract}
\newpage

\section{Introduction}

In this paper, we study secure communications in multi-user
interference networks from an information-theoretic point of view. The
security of communication was first studied by Shannon via a noiseless
wiretap channel \cite{Shannon:1949}. Noisy wiretap channel was
introduced by Wyner who determined its capacity-equivocation region
for the degraded case \cite{wyner}. His result was generalized to
arbitrary, not necessarily degraded, wiretap channels by Csiszar and
Korner \cite{csiszar}, and extended to Gaussian wiretap channels by
Leung-Yan-Cheong and Hellman \cite{gaussian}. This line of research
has been subsequently extended to many multi-user settings, e.g., broadcast channels with confidential messages \cite{secrecy_ic, xu_bounds_bc_cm_it_09}, multi-receiver wiretap channels \cite{fading1, bagherikaram_bc_2008, ersen_bc_asilomar_08, ersen_eurasip_2009} (see also a survey on extensions of these to MIMO channels \cite{ersen_jcn_2010}), interference channels with confidential messages \cite{secrecy_ic, he_outerbound_gic_cm_ciss_09}, interference channels with external eavesdroppers \cite{koyluoglu_ic_external}, multiple access wiretap channels \cite{tekin_gmac_w, cooperative_jamming, ersen_mac_allerton_08, liang_mac_cm_08 ,ersen_mac_book_chapter}, wiretap channels with helpers \cite{wiretap_channel_with_one_helper}, relay eavesdropper channels \cite{relay_1, relay_2, relay_3, relay_4, he_untrusted_relay, ersen_crbc_2011}, compound wiretap channels \cite{compound_wiretap_channel, ersen_ulukus_degraded_compound}, etc. While the channel models involving a single transmitter, such as broadcast channels with confidential messages and multi-receiver wiretap channels, are relatively better understood, the channel models involving multiple independent transmitters, such as interference channels with confidential messages and/or external eavesdroppers, multiple access wiretap channels, wiretap channels with helpers, and relay-eavesdropper channels, are much less understood. The exact secrecy capacity regions of all these multiple-transmitter models remain unknown, even in the case of simple Gaussian channels. In the absence of exact secrecy capacity regions, achievable
secure degrees of freedom (d.o.f.) at high signal-to-noise ratio (SNR) regimes
has been studied in the literature \cite{he_k_gic_cm_09, koyluoglu_k_user_gic_secrecy, xie_k_user_ia_compound, secrecy_ia_new, xiang_he_thesis, secrecy_ia5,raef_mac_it_12, secrecy_ia1, interference_alignment_compound_channel, xie_gwch_allerton, xie_sdof_networks_in_prepare, xie-ulukus-ciss13, xie_layered_network_journal}. In this paper, we focus on the
$K$-user interference channel with secrecy constraints, and determine its exact sum secure d.o.f.

\begin{figure}[t]
\centering
\includegraphics[scale=0.9]{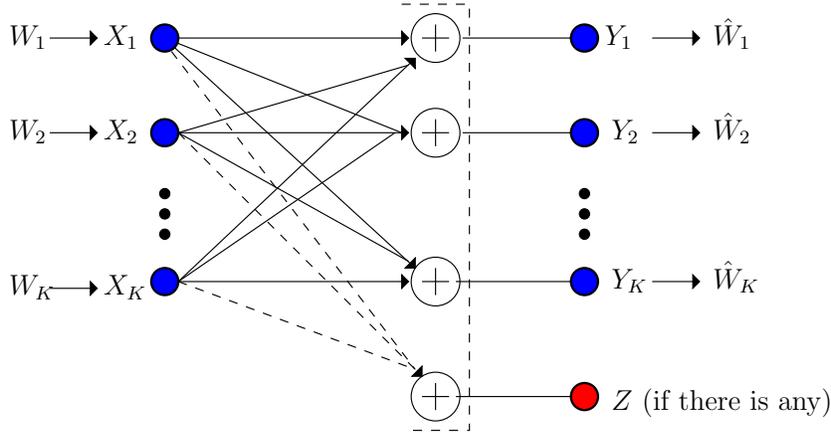}
\caption{$K$-user Gaussian interference channel with secrecy constraints.}
\label{fig:kic-general}
\end{figure}

\begin{figure}[t]
\centerline{\begin{tabular}{ccc}
\subfigure{\includegraphics[scale=0.7]{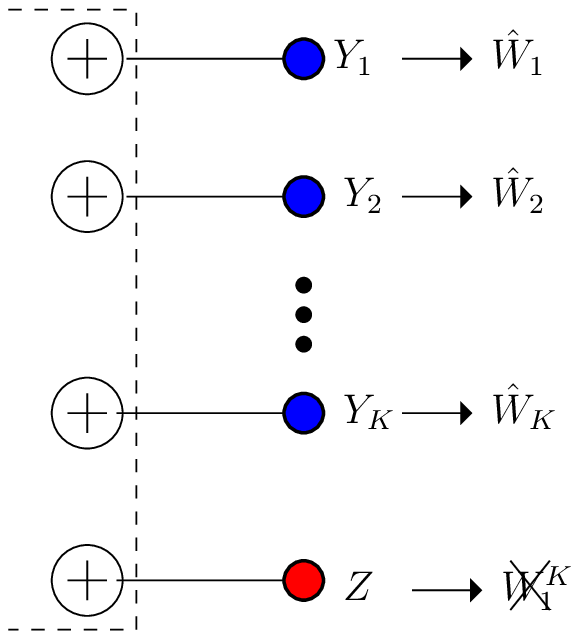}} \hspace*{0.4in}&
\subfigure{\includegraphics[scale=0.7]{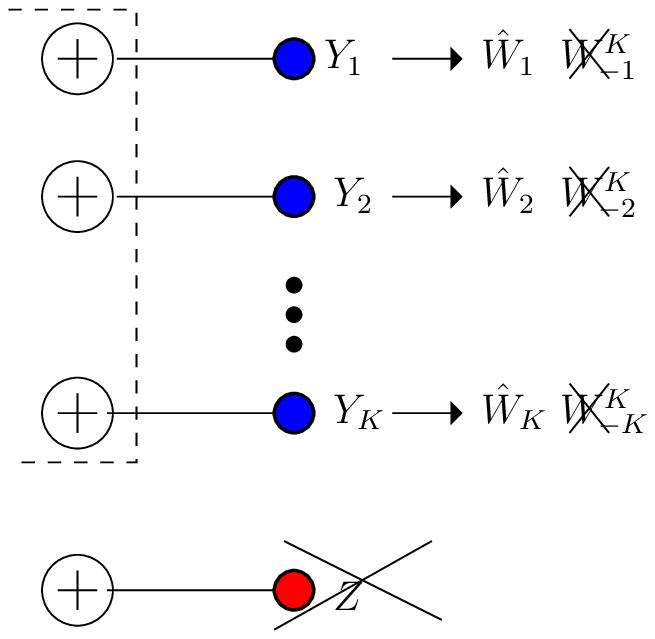}} \hspace*{0.4in}&
\subfigure{\includegraphics[scale=0.7]{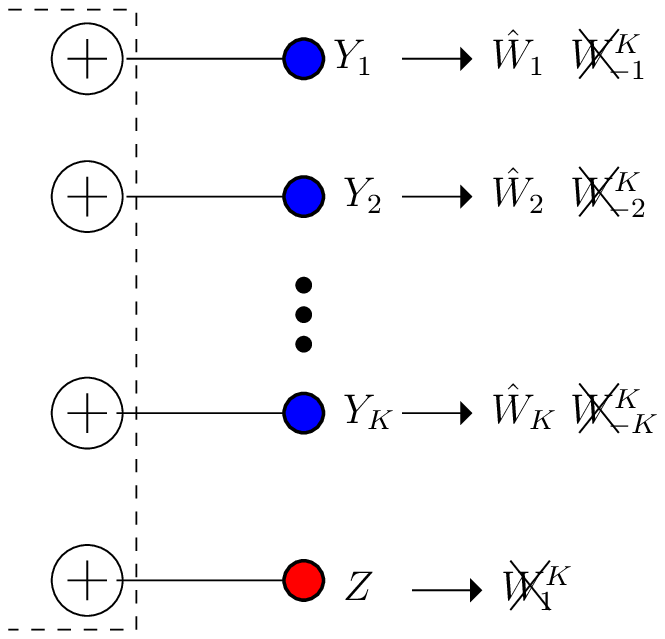}}\\
(a) & (b) & (c)
\end{tabular}}
\caption{The receiver sides of the three channel models: (a) $K$-user
  IC-EE, (b) $K$-user IC-CM, and
  (c) $K$-user IC-CM-EE, where $W_{-i}^K \defn
  \{W_1,\ldots,W_{i-1},W_{i+1},\ldots, W_K\}$.}
\label{fig:kic-subfigs}
\end{figure}

The $K$-user Gaussian interference channel with secrecy constraints consists of $K$ \linebreak transmitter-receiver pairs each wishing to have secure communication over a Gaussian interference channel (IC); see Figure~\ref{fig:kic-general}. We consider three different secrecy constraints: 1) $K$-user interference channel with one external eavesdropper (IC-EE), where $K$ transmitter-receiver pairs wish to have secure communication against an external eavesdropper, see Figure~\ref{fig:kic-subfigs}(a). 2) $K$-user interference channel with confidential messages (IC-CM), where there are no external eavesdroppers, but each transmitter-receiver pair wishes to secure its communication against the remaining $K-1$ receivers, see Figure~\ref{fig:kic-subfigs}(b). 3) $K$-user interference channel with confidential messages and one external eavesdropper (IC-CM-EE), which is a combination of the previous two cases, where each transmitter-receiver pair wishes to secure its communication against the remaining $K-1$ receivers and the external eavesdropper, see Figure~\ref{fig:kic-subfigs}(c).

In the Gaussian wiretap channel, the secrecy capacity is the
difference between the channel capacities of the transmitter-receiver
and the transmitter-eavesdropper pairs \cite{gaussian}. It is well-known that this
difference does not scale with the SNR, and hence the secure d.o.f.~of
the Gaussian wiretap channel is zero, indicating a severe penalty due
to secrecy in this case. Fortunately, this does not hold in most multi-user
scenarios, including the interference channel. Reference
\cite{he_k_gic_cm_09} showed that nested lattice codes and layered
coding are useful in providing positive sum secure d.o.f. for the
$K$-user IC-CM; their result gave a sum secure d.o.f.~of less than
$\frac{3}{4}$ for $K=3$.  Reference
\cite{koyluoglu_k_user_gic_secrecy} used interference alignment to
achieve a sum secure d.o.f.~of $\frac{K(K-2)}{2K-2}$ for the $K$-user
IC-CM, which gave $\frac{3}{4}$ for $K=3$. Based on the same idea,
\cite{xie_k_user_ia_compound, koyluoglu_k_user_gic_secrecy} achieved a
sum secure d.o.f. of $\frac{K(K-1)}{2K}$ for the $K$-user IC-EE, which
gave $1$ for $K=3$. The approach used in \cite{xie_k_user_ia_compound,
koyluoglu_k_user_gic_secrecy} is basically to evaluate the secrecy
performance of the interference alignment technique
\cite{interference_alignment} devised originally for the $K$-user interference
channel without any secrecy constraints. Since the original interference alignment scheme puts all
of the interfering signals into the same reduced-dimensionality sub-space at a receiver, it
naturally provides a certain amount of secrecy to those signals
as an unintended byproduct, because the interference signals in
this sub-space create uncertainty for one another and make it
difficult for the receiver to decode them. However, since the end-goal of \cite{interference_alignment} is \emph{only} to achieve reliable decoding of the transmitted messages at their intended receivers, the d.o.f. it provides is sub-optimal when \emph{both} secrecy and reliability of messages are considered.

Recently, the \emph{exact} sum secure d.o.f. of the two-user IC-CM was
obtained to be $\frac{2}{3}$ in \cite{xie_sdof_networks_in_prepare}.
This reference showed that while interference alignment is a key
ingredient in achieving positive secure d.o.f., a more intricate
design of the signals is needed to achieve the simultaneous end-goals
of reliability at the desired receivers and secrecy at the
eavesdroppers. In particular, in \cite{xie_sdof_networks_in_prepare},
each transmitter sends both message carrying signals, as well as
cooperative jamming signals. This random mapping of the message
carrying signals to the channel inputs via cooperative jamming signals
may be interpreted as channel prefixing \cite{csiszar}. Both the
message carrying signals and the cooperative jamming signals come from
the same discrete alphabet, and hence are structured. In addition, the
signals are carefully aligned at the legitimate receivers and the
eavesdroppers using real interference alignment \cite{real_inter_align}.
In particular, at each receiver, the unintended message and both jamming signals are
constrained in the same interference sub-space, providing an
interference-free sub-space for the intended message. Further, inside the interference sub-space,
each unintended message is protected by aligning it with
the jamming signal from the other transmitter. Such a perfect alignment provides a constant upper bound for the information leakage rate.

In this paper, we generalize the results in
\cite{xie_sdof_networks_in_prepare} to the case of $K$-user
interference channel, for $K>2$. Our generalization has three main
components:

\begin{enumerate}
\item While \cite{xie_sdof_networks_in_prepare} considered IC-CM only, we
consider both IC-CM and IC-EE and their combination IC-CM-EE in a
unified framework.  To this end, we show converses separately for
IC-EE and IC-CM, which imply a converse for IC-CM-EE; and we show
achievability for IC-CM-EE, which implies achievability
for IC-EE and IC-CM. The achievability and converse meet giving an
{\it exact} sum secure d.o.f.~of $\frac{K(K-1)}{2K-1}$ for all three models.

\item For achievability: In the case of two-user IC-CM in \cite{xie_sdof_networks_in_prepare},
each message needs to be delivered reliably to one receiver and needs to be
protected from another receiver. This requires alignment at two receivers, which is achieved
in \cite{xie_sdof_networks_in_prepare} by simply choosing transmission coefficients properly, which cannot be extended to the $K$-user case here. In the $K$-user
IC-CM-EE case, we need to deliver each message to a receiver, while
protecting it from $K$ other receivers. This requires designing
signals in order to achieve alignment at $K+1$ receivers
simultaneously: at one receiver (desired receiver) we need alignment to ensure that the
largest space is made available to message carrying signals for their
reliable decodability, and at $K$ other receivers, we need to align
cooperative jamming signals with message carrying signals to protect
them. These requirements create two challenges: i) aligning multiple signals
simultaneously at multiple receivers, and ii) upper bounding
the information leakage rates by suitable functions which can be made small.
We overcome these challenges by using an asymptotical approach \cite{real_inter_align_exploit},
where we introduce many signals that carry each message and align them simultaneously at multiple receivers only order-wise (i.e., align most of them, but not all of them), and by developing a method
to upper bound the information leakage rate by a function which can be made small.
In contrast to the constant upper bound for the information leakage rate in \cite{xie_sdof_networks_in_prepare}, here the upper bound is not constant, but a function which can be made small. This is due to the non-perfect (i.e., only asymptotical) alignment.

\item For the converse: To the best of our knowledge, the only known
upper bound for the sum secure \dof of the $K$-user interference
channel with secrecy constraints is $\frac{K}{2}$, which is the upper
bound with no secrecy constraints \cite{interference_alignment}. The
upper bounding technique for the two-user IC-CM in
\cite{xie_sdof_networks_in_prepare} considers one single confidential
message against the corresponding unintended receiver each time, since
in that case the eavesdropping relationship is straightforward: for each
message there is only one eavesdropper and for each
eavesdropper there is only one confidential message. However, in the
case of $K$-user IC, each message is required to be kept secret against
multiple eavesdroppers and each eavesdropper is associated
with multiple unintended messages. To develop a tight converse, we focus
on the eavesdropper as opposed to the message. In the converse for IC-EE,
we consider the sum rate of all of the messages
eavesdropped by the external eavesdropper. We sequentially apply the {\it role
of a helper lemma} in \cite{xie_sdof_networks_in_prepare} to each
transmitter by treating its signal as a helper to another specific transmitter. In the
converse for IC-CM, for each receiver (which also is an eavesdropper), we consider the sum rate of
all unintended messages, and again apply the {\it role of a helper lemma} in a specific structure.
\end{enumerate}

\section{System Model, Definitions and the Result}
\label{kic-model-ee}

The input-output relationships for a $K$-user Gaussian interference channel
with secrecy constraints (Figure~\ref{fig:kic-general}) are given by
\begin{align}
  \label{eqn:kic-channel-model-ee-1}
  Y_i & = \sum_{j=1}^K h_{ji} X_j + N_i, \qquad i =1,\ldots,K \\
  \label{eqn:kic-channel-model-ee-2}
  Z   & = \sum_{j=1}^K g_{j} X_j + N_Z
\end{align}
where $Y_i$ is the channel output of receiver $i$, $Z$ is the channel output of the external eavesdropper (if there is any), $X_i$ is the channel input of transmitter $i$, $h_{ji}$ is the channel gain of the $j$th transmitter to the $i$th receiver, $g_j$ is the channel gain of the $j$th transmitter to the eavesdropper (if there is any), and $\{N_1,\ldots,N_K,N_Z\}$ are mutually independent zero-mean unit-variance Gaussian random variables. All the channel gains are time-invariant, and independently drawn from continuous distributions. We further assume that all $h_{ji}$ are non-zero, and all $g_j$ are non-zero if there is an external eavesdropper. All channel inputs satisfy average power constraints, $\E\left[X^2_{i}\right] \le
P$, for $i=1,\ldots, K$.

Each transmitter $i$ intends to send a message $W_i$, uniformly chosen from a set $\mathcal{W}_i$, to receiver $i$. The rate of the message is $R_i\defn\frac{1}{n}\log|\mathcal{W}_i|$, where $n$ is the number of channel uses. Transmitter $i$ uses a stochastic function $f_i: \mathcal{W}_i\to \bfX_i$ to encode the message, where $\bfX_i\defn X_i^n$ is the $n$-length channel input of user $i$. We
use boldface letters to denote $n$-length vector signals, e.g., $\bfX_i\defn X_i^n$, $\bfY_j\defn Y_j^n$, $\bfZ\defn Z^n$, etc. The legitimate receiver $j$ decodes the message as $\hat{W}_j$ based
on its observation $\mathbf{Y}_j $. A rate tuple $(R_1,\ldots,R_K)$ is said to be achievable if for any $\epsilon>0$, there exist joint $n$-length codes such that each receiver $j$ can decode the corresponding message reliably, i.e., the probability of decoding error is less than $\epsilon$ for all messages,
\begin{equation}
 \max_{j}\pr\left[W_j\neq\hat{W}_j\right] \le \epsilon
\end{equation}
and the corresponding secrecy requirement is satisfied. We consider three different secrecy requirements:
\begin{enumerate}
\item[{1)}] In IC-EE, Figure~\ref{fig:kic-subfigs}(a), all of the messages are kept information-theoretically secure against the external eavesdropper,
\begin{align}
  \label{eqn:kic-secrecy-constraint-cmee-1}
  H(W_1,\ldots, W_K|\bfZ) & \ge H(W_1,\ldots, W_K) - n \epsilon
\end{align}
\item[{2)}] In IC-CM, Figure~\ref{fig:kic-subfigs}(b), all unintended messages are kept
  information-theoretically secure against each receiver,
\begin{align}
  \label{eqn:kic-secrecy-constraint-cmee-2}
  H(W_{-i}^K|\bfY_i) & \ge H(W_{-i}^K) - n \epsilon,
  \qquad i=1,\ldots,K
\end{align}
where $W_{-i}^K \defn \{W_1,\ldots,W_{i-1},W_{i+1},\ldots, W_K\}$.
\item[{3)}] In IC-CM-EE, Figure~\ref{fig:kic-subfigs}(c), all of the messages are kept information-theoretically secure against both the $K-1$ unintended receivers and the eavesdropper, i.e., we impose both secrecy constraints in \eqn{eqn:kic-secrecy-constraint-cmee-1} and
  \eqn{eqn:kic-secrecy-constraint-cmee-2}.
\end{enumerate}

The supremum of all sum achievable secrecy rates is the sum secrecy capacity $C_{s,\Sigma}$, and the sum secure d.o.f., $D_{s,\Sigma}$, is defined as
\begin{equation}
  D_{s,\Sigma} \defn \lim_{P\to\infty}  \frac{C_{s,\Sigma}}{\frac{1}{2}\log P}
  = \lim_{P\to\infty}  \sup \frac{R_1+\cdots+R_K}{\frac{1}{2}\log P}
\label{eqn:kic-ee-sec-dof-defn}
\end{equation}
The main result of this paper is stated in the following theorem.

\begin{theorem}
  \label{eqn:kic-ds-final}
The sum secure d.o.f. of the $K$-user IC-EE, IC-CM, and IC-CM-EE is $\frac{K(K-1)}{2K-1}$ for almost all channel gains.
\end{theorem}

\section{Preliminaries}

\subsection{ Role of a Helper Lemma}
For completeness, we repeat Lemma~2 in \cite{xie_sdof_networks_in_prepare} here, which is called {\it role of a helper lemma}. This lemma identifies a constraint on the signal of a given transmitter, based on the decodability of another transmitter's message at its intended receiver.

\begin{lemma}[\!\! \cite{xie_sdof_networks_in_prepare}]
\label{lemma:gwch_general_ub_for_helper}
For reliable decoding of the $k$th transmitter's signal at the $k$th receiver, the
channel input of transmitter $i\neq k$, $\bfX_i$, must satisfy
\begin{equation}
 h(\bfX_i + \tilde\bfN)\le  h(\bfY_k)  - n R_k + n {c}
  \label{eqn:gwch_general_ub_for_helper}
\end{equation}
where $c$ is a constant which does not depend on $P$, and $\tN$ is a new
Gaussian random variable independent of all other random variables with $\sigma_{\tN}^2 <
\frac{1}{h_{ik}^2}$, and $\tilde\bfN$ is an i.i.d.~sequence of $\tilde N$.
\end{lemma}

Lemma~\ref{lemma:gwch_general_ub_for_helper} gives an upper bound on the
differential entropy of (a noisy version of) the signal of any given transmitter,
transmitter $i$ in \eqn{eqn:gwch_general_ub_for_helper}, in terms of the differential entropy of
the channel output and the message rate $n R_k = H(W_k)$, of a user $k$, based on the decodability of message $W_k$ at its intended receiver.
The inequality in this lemma, \eqn{eqn:gwch_general_ub_for_helper}, can
alternatively be interpreted as an upper bound on the message rate, i.e., on
$n R_k$, in terms of the difference of the differential entropies of the channel
output of a receiver $k$ and the channel input of a transmitter $i$; in
particular, the higher the differential entropy of the signal coming from user $i$, the lower this upper bound will be on the rate of user $k$. This motivates not using i.i.d.~Gaussian
signals which have the highest differential entropy. Also note that this lemma does not
involve any secrecy constraints, and is based only on the decodability of the messages at their
intended receivers.

\subsection{Real Interference Alignment}

\subsubsection{Pulse Amplitude Modulation}
For a point-to-point scalar Gaussian channel,
\begin{equation}
Y = X + Z
\end{equation}
with additive Gaussian noise $Z \sim \mathcal{N}(0,\sigma^2)$ and
an input power constraint $\mathe{X^2} \le P$, assume that
the input symbols are drawn from a PAM constellation,
\begin{equation}
C(a,Q) = a \{ -Q, -Q+1, \ldots, Q-1,Q\}
\label{constel}
\end{equation}
where $Q$ is a positive integer and $a$ is a real number to normalize
the transmit power. Note that, $a$ is also the minimum distance
$d_{min}(C)$ of this constellation, which has the probability
of error
\begin{equation}
\pe(e) \le \exp\left(  - \frac{d_{min}^2}{8 \sigma^2}\right)  = \exp\left(  - \frac{a^2}{8 \sigma^2}\right)
\end{equation}
The transmission rate of this PAM scheme is
\begin{equation}
R =  \log( 2 Q + 1)
\end{equation}
since there are $2Q+1$ signalling points in the constellation. For any small enough $\delta>0$, if we choose $Q = P^{\frac{1-\delta}{2}}$ and $a=\gamma P^{\frac{\delta}{2}}$, where
$\gamma$ is a constant to normalize the transmit power, which is
independent of $P$, then
\begin{equation}
\pe(e)
\le \exp\left( -\frac{\gamma^2 P^{{ \delta}}}{8\sigma^2} \right)
\qquad \hbox{and} \qquad
R
 \ge \frac{1-\delta}{2} \log P
\end{equation}
and we can have $\pe(e)\to 0$ and $R\to\frac{1}{2}\log P$ as
$P\to\infty$. That is, we can have reliable communication at rates
approaching $\frac{1}{2}\log P$, and therefore have $1$ d.o.f.

\subsubsection{Real Interference Alignment}

This PAM scheme for the point-to-point scalar channel can be
generalized to multiple data streams.  Let the transmit signal be
\begin{equation}
x = \mb{a}^T \mb{b} = \sum^L_{i=1} a_i b_i
\end{equation}
where $a_1,\ldots, a_L$ are rationally independent real
numbers\footnote{ $a_1, \ldots, a_L$ are rationally independent if
whenever $q_1,\ldots,q_L$ are rational numbers  then
$\sum^L_{i=1} q_i a_i =0$  implies $q_i=0$ for all $i$.  } and  each
$b_i$ is drawn independently from the constellation $C(a,Q)$ in (\ref{constel}).
The real value $x$ is a combination of  $L$ data streams, and the constellation observed at
the receiver consists of $(2 Q+1)^L$ signal points.

By using the Khintchine-Groshev theorem of Diophantine approximation
in number theory, \cite{real_inter_align_exploit,real_inter_align}
bounded the minimum distance $d_{min}$ of points in the receiver's
constellation: For any $\delta>0$, there exists a
constant $k_\delta$, such that
\begin{equation}
\label{ria:lb_of_d}
d_{min} \ge \frac{ k_\delta  a}{Q^{L-1+\delta}}
\end{equation}
for almost all rationally independent $\{a_i\}_{i=1}^L$, except for a set
of Lebesgue measure zero. Since the minimum distance of the receiver
constellation is lower bounded, with proper choice of $a$ and $Q$, the
probability of error can be made arbitrarily small, with rate $R$ approaching
$\frac{1}{2} \log P$.  This result is stated in the following lemma.

\begin{lemma}[\!\! \cite{real_inter_align_exploit,real_inter_align}]
\label{lemma:ria_real_alignment}
For any small enough $\delta>0$, there exists a positive constant
$\gamma$, which is independent of $P$, such that if we choose
\begin{equation}
Q = P^{\frac{1-\delta}{2(L+\delta)}}
\qquad \mbox{and} \qquad
a=\gamma \frac{P^{\frac{1}{2}}}{Q}
\end{equation}
then the average power constraint is satisfied, i.e., $\mathe{X^2}\le P $,
and for almost all $\{a_i\}_{i=1}^L$, except for a set of Lebesgue measure zero,
the probability of error is bounded by
\begin{equation}
  \mathrm{Pr}(e)
\le
\exp\left( - \eta_\gamma P^{{ \delta}} \right)
\end{equation}
where $\eta_\gamma$ is a positive constant which is
independent of $P$.
\end{lemma}

Furthermore, as a simple extension, if $b_i$ are sampled independently
from different constellations $C_i(a,Q_i)$, the lower bound in
\eqn{ria:lb_of_d} can be modified as
\begin{equation}
d_{min} \ge \frac{ k_\epsilon  a}{(\max_i Q_i)^{L-1+\epsilon}}
\end{equation}

\section{Converse for IC-EE}

In this section, we develop a converse for the $K$-user IC-EE (see
Figure~\ref{fig:kic-subfigs}(a)) defined
in \eqn{eqn:kic-channel-model-ee-1} and
\eqn{eqn:kic-channel-model-ee-2} with the secrecy constraint \eqn{eqn:kic-secrecy-constraint-cmee-1}.  We start with the sum rate:
\begin{align}
n\sum_{i=1}^K R_i
& = \sum_{i=1}^K H(W_i) = H(W_1^K)\\
& \le I(W_1^K;\bfY_1^K)  - I(W_1^K;\bfZ) + n \nextscnu \\
& \le I(W_1^K;\bfY_1^K,\bfZ)  - I(W_1^K;\bfZ) + n \nextscnu \\
& = I(W_1^K;\bfY_1^K|\bfZ)  + n \nextscnu \\
& \le I(\bfX_1^K;\bfY_1^K|\bfZ)  + n \nextscnu \\
& = h(\bfY_1^K|\bfZ) - h(\bfY_1^K|\bfZ,\bfX_1^K)  + n \nextscnu \\
& = h(\bfY_1^K|\bfZ) - h(\bfN_1^K|\bfZ,\bfX_1^K)  + n \nextscnu \\
& \le h(\bfY_1^K|\bfZ)  + n \nextsc\\
& = h(\bfY_1^K,\bfZ)  -h(\bfZ) + n \nextscnu
\label{kid:converse-ic-ee-continue-from}
\end{align}
where $W_1^K\defn \{W_j\}_{j=1}^K$, $\bfX_1^K\defn \{\bfX_j\}_{j=1}^K$,
$\bfY_1^K\defn \{\bfY_j\}_{j=1}^K$, and all the  $c_i$s in this paper are constants which do
not depend on $P$.

For each $j$, we introduce $\tilde\bfX_j = \bfX_j +
\tilde\bfN_j$, where $\tilde\bfN_j$ is an i.i.d.~sequence of $\tN_j$ which is a
zero-mean Gaussian random variable with variance $\sigma_j^2 < \min(\min_i 1/h_{ji}^2,1/g_j^2)$.
Also, $\{\tN_j\}_{j=1}^K$ are mutually independent, and are independent of all
other random variables. Continuing from \eqn{kid:converse-ic-ee-continue-from},
\begin{align}
n\sum_{i=1}^K R_i
& \le h(\tilde\bfX_1^K, \bfY_1^K,\bfZ)- h(\tilde\bfX_1^K|
\bfY_1^K,\bfZ)-h(\bfZ) + n \nextscnu \\
& \le h(\tilde\bfX_1^K, \bfY_1^K,\bfZ)- h(\tilde\bfX_1^K|
\bfX_1^K,\bfY_1^K,\bfZ)-h(\bfZ) + n \nextscnu \\
& = h(\tilde\bfX_1^K, \bfY_1^K,\bfZ)- h(\tilde\bfN_1^K)-h(\bfZ) + n \nextscnu \\
& \le h(\tilde\bfX_1^K, \bfY_1^K,\bfZ)-h(\bfZ) + n \nextsc  \\
& = h(\tilde\bfX_1^K) + h( \bfY_1^K,\bfZ|\tilde\bfX_1^K)-h(\bfZ) + n \nextscnu
\\
& \le h(\tilde\bfX_1^K) -h(\bfZ) + n \nextsc \label{into-that}
\end{align}
where $\tilde\bfX_1^K\defn \{\tilde\bfX_j\}_{j=1}^K$, and the last inequality is due to the fact that $h(\bfY_1^K,\bfZ|\tilde\bfX_1^K)\leq nc'$, i.e., given all the channel inputs (disturbed by small Gaussian noises), the channel outputs can be
\emph{reconstructed}, which is shown as follows
\begin{align}
& h(\bfY_1^K,\bfZ|\tilde\bfX_1^K) \nl
&\quad\quad \le
  \left[\sum_{j=1}^K h(\bfY_j|\tilde\bfX_1^K) \right]
  + h(\bfZ|\tilde\bfX_1^K)
  \\
&\quad\quad =
  \left[ \sum_{j=1}^K h\left(\sum_{i=1}^{K} h_{ij} (\tilde\bfX_i - \tilde\bfN_i) + \bfN_j \Bigg|
  \tilde\bfX_1^K\right)  \right]
     + h\left(\sum_{i=1}^{K} g_i (\tilde\bfX_i - \tilde\bfN_i) + \bfN_Z \Bigg|
   \tilde\bfX_1^K\right)
  \\
&\quad\quad =
  \left[ \sum_{j=1}^K h\left(- \sum_{i=1}^{K} h_{ij}   \tilde\bfN_i + \bfN_j \Bigg|
  \tilde\bfX_1^K\right)  \right]
     + h\left(- \sum_{i=1}^{K} g_i \tilde\bfN_i + \bfN_Z \Bigg|
   \tilde\bfX_1^K\right)
  \\
&\quad\quad \le
  \left[ \sum_{j=1}^K h\left(- \sum_{i=1}^{K} h_{ij}   \tilde\bfN_i + \bfN_j \right)  \right]
   + h\left(- \sum_{i=1}^{K} g_i \tilde\bfN_i + \bfN_Z \right)
  \\
&\quad\quad \stackrel{\triangle}{=} n \nextsc
         \label{eqn:kic_reconstruction}
\end{align}

Next, we note
\begin{equation}
h(\tilde\bfX_j)
\le h( g_j \bfX_j + \bfN_Z) + n \nextsc
\le h(\bfZ) + n \nextscnu, \qquad  j=1,\ldots,K \label{inserting-this}
\end{equation}
where the inequalities are due to the differential entropy version of \cite[Problem
2.14]{cover_it_book}. Inserting (\ref{inserting-this}) into (\ref{into-that}), for any $j=1,\ldots,K$, we get
\begin{align}
n\sum_{i=1}^K R_i
& \le h(\tilde\bfX_1^K) -h(\bfZ) + n c_3 \\
& \le \sum_{i=1}^K h(\tilde\bfX_i) -h(\bfZ) + n c_3\\
& \le \sum_{i=1,i\neq j}^K h(\tilde\bfX_i) + n \nextsc \label{apply-rohl}
\end{align}
which means that the net effect of the presence of an eavesdropper is to \emph{eliminate} one of the channel inputs; we call this the \emph{secrecy penalty}.

We apply the {\it role of a helper lemma}, Lemma~\ref{lemma:gwch_general_ub_for_helper}, to  each $\tilde\bfX_i$ with $k=i+1$ (for $i=K$, $k=1$), in (\ref{apply-rohl}) as
\begin{align}
n\sum_{i=1}^K R_i
& \le h(\tilde\bfX_1) +  h(\tilde\bfX_2) + \cdots
+ h(\tilde\bfX_{j-1}) +h(\tilde\bfX_{j+1}) + \cdots + h(\tilde\bfX_K) + n\nextsc
\\
& \le \left[ h(\bfY_2) - n R_2 \right]
+ \left[ h(\bfY_3) - n R_3 \right] + \cdots
 + \left[h(\bfY_{j}) - n R_{j} \right]\nl
&\quad+ \left[ h(\bfY_{j+2}) - n R_{j+2} \right]
+ \cdots +\left[ h(\bfY_{K}) - n R_{K} \right] + \left[ h(\bfY_1) - n R_1 \right]
+ n \nextsc
\end{align}
By noting that $h(\bfY_i) \le \frac{n}{2}\log P + n c_i'$
for each $i$, we have
\begin{align}
2 n\sum_{i=1}^K R_i \le (K-1) \left( \frac{n}{2}\log P \right) + n R_{{(j+1)} \bmod{K}} + n \nextsc
\label{eqn:kic-ee-ub-general-j}
\end{align}
for $j=1,\ldots,K$. Therefore, we have a total of $K$ bounds in \eqn{eqn:kic-ee-ub-general-j} for $j=1,\ldots,K$. Summing these $K$ bounds, we obtain:
\begin{align}
(2 K - 1) n\sum_{i=1}^K R_i \le K (K-1) \left( \frac{n}{2}\log P \right) + n \nextsc
\end{align}
which gives
\begin{align}
  D_{s,\Sigma} \le \frac{ K (K-1)}{2 K - 1}
\end{align}
completing the converse for IC-EE.

\section{Converse for IC-CM}
In this section, we develop a converse for the $K$-user IC-CM
(see Figure~\ref{fig:kic-subfigs}(b)). We
focus on the secrecy constraint
\eqn{eqn:kic-secrecy-constraint-cmee-2} at a single receiver, say $j$,
as an eavesdropper, and start with the sum rate corresponding to all
unintended messages at receiver $j$:
\begin{align}
n\sum_{i=1,i\neq j}^K R_i
& = \sum_{i=1,i\neq j}^K H(W_i) = H(W_{-j}^K)\\
& \le I(W_{-j}^K;\bfY_{-j}^K)  - I(W_{-j}^K;\bfY_j) + n \nextsc \\
& \le I(W_{-j}^K;\bfY_{-j}^K,\bfY_j)  - I(W_{-j}^K;\bfY_j) + n \nextscnu \\
& =   I(W_{-j}^K;\bfY_{-j}^K|\bfY_j)  + n \nextscnu \\
& \le I(\bfX_{-j}^K;\bfY_{-j}^K|\bfY_j)  + n \nextscnu \\
& =   h(\bfY_{-j}^K|\bfY_j) - h(\bfY_{-j}^K|\bfY_j,\bfX_{-j}^K)  + n \nextscnu \\
& \le h(\bfY_{-j}^K|\bfY_j) - h(\bfY_{-j}^K|\bfY_j,\bfX_1^K)  + n \nextscnu \\
& =   h(\bfY_{-j}^K|\bfY_j) - h(\bfN_{-j}^K|\bfY_j,\bfX_1^K)  + n \nextscnu \\
& \le h(\bfY_{-j}^K|\bfY_j)  + n \nextsc \\
& =   h(\bfY_{-j}^K,\bfY_j) - h(\bfY_j)  + n \nextscnu \\
& =   h(\bfY_1^K)-h(\bfY_j) + n \nextscnu
\label{kic:converse-ic-cm-continue-from2}
\end{align}
where $W_{-j}^K \defn \{W_i\}_{i=1,i\neq j}^K$ is the message
set containing all unintended messages with respect to receiver $j$,
$\bfX_{-j}^K\defn \{\bfX_i\}_{i=1,i\neq j}^K$
and $\bfY_{-j}^K\defn \{\bfY_i\}_{i=1,i\neq j}^K$.

For each $j$, we introduce $\tilde\bfX_j = \bfX_j +
\tilde\bfN_j$, where $\tilde\bfN_j$ is an i.i.d.~sequence of $\tN_j$ which is a
zero-mean Gaussian random variable with variance $\sigma_j^2 < \min_i 1/h_{ji}^2$.
Also, $\{\tN_j\}_{j=1}^K$ are mutually independent, and are independent of all
other random variables. Continuing from \eqn{kic:converse-ic-cm-continue-from2},
\begin{align}
n\sum_{i=1,i\neq j}^K R_i
& \le h(\tilde\bfX_1^K, \bfY_1^K)- h(\tilde\bfX_1^K| \bfY_1^K)-h(\bfY_j) + n
\nextscnu \\
& \le h(\tilde\bfX_1^K, \bfY_1^K)- h(\tilde\bfX_1^K| \bfY_1^K,\bfX_1^K)-h(\bfY_j) + n
\nextscnu \\
& =   h(\tilde\bfX_1^K, \bfY_1^K)- h(\tilde\bfN_1^K)-h(\bfY_j) + n
\nextscnu \\
& \le h(\tilde\bfX_1^K, \bfY_1^K)-h(\bfY_j) + n
\nextsc \\
& =   h(\tilde\bfX_1^K) + h(\bfY_1^K|\tilde\bfX_1^K)-h(\bfY_j) + n
\nextscnu \\
& \le h(\tilde\bfX_1^K) -h(\bfY_j) + n \nextsc \label{follow-from-here}
\end{align}
where the last inequality is due to the fact that $h(\bfY_1^K|\tilde\bfX_1^K)\leq n c'$, i.e., given all the channel inputs (disturbed by small Gaussian noises), the channel outputs can be
\emph{reconstructed}, which is shown as follows
\begin{align}
 h(\bfY_1^K|\tilde\bfX_1^K)
& \le
  \sum_{j=1}^K h(\bfY_j|\tilde\bfX_1^K)
  \\
& =
  \sum_{j=1}^K h\left(\sum_{i=1}^{K} h_{ij} (\tilde\bfX_i - \tilde\bfN_i) + \bfN_j \Bigg|
  \tilde\bfX_1^K\right)    \\
& =
   \sum_{j=1}^K h\left(- \sum_{i=1}^{K} h_{ij}   \tilde\bfN_i + \bfN_j \Bigg|
  \tilde\bfX_1^K\right)
  \\
& \le
  \sum_{j=1}^K h\left(- \sum_{i=1}^{K} h_{ij}   \tilde\bfN_i + \bfN_j \right)   \\
& \stackrel{\triangle}{=} n \nextsc
\end{align}

We apply the {\it role of a helper lemma}, Lemma~\ref{lemma:gwch_general_ub_for_helper}, to  each $\tilde\bfX_i$ with $k=i+1$ (for $i=K$, $k=1$), in (\ref{follow-from-here}) as
\begin{align}
n\sum_{i=1,i\neq j}^K R_i
& \le h(\tilde\bfX_1^K) -h(\bfY_j) + n c_{14} \\
& \le \sum_{i=1}^K h(\tilde\bfX_i) -h(\bfY_j) + n c_{14} \\
& \le \sum_{i=1}^{K-1} \Big[ h(\bfY_{i+1}) - n R_{i+1} \Big] + \Big[ h(\bfY_{1}) - n R_{1} \Big] -h(\bfY_j) + n \nextsc \\
& =   \sum_{i=1}^K \Big[ h(\bfY_{i}) - n R_{i} \Big] -h(\bfY_j) + n
\nextscnu
\end{align}
By noting that $h(\bfY_i) \le \frac{n}{2}\log P + n c_i'$
for each $i$, we have
\begin{align}
n R_j + 2 n\sum_{i=1,i\neq j}^K R_i
& \le  \sum_{i=1,i\neq j}^K  h(\bfY_{i})  + n
\nextscnu  \\
& \le  (K-1) \left(\frac{n}{2}\log P \right) + n
\nextsc
\label{eqn:kic-cm-ub-general-j}
\end{align}
for $j=1,\ldots,K$. Therefore, we have a total of $K$ bounds in \eqn{eqn:kic-cm-ub-general-j} for
$j=1,\ldots,K$. Summing these $K$ bounds, we obtain:
\begin{align}
(2 K - 1) n\sum_{i=1}^K R_i \le K (K-1) \left( \frac{n}{2}\log P \right) + n \nextsc
\end{align}
which gives
\begin{align}
  D_{s,\Sigma} \le \frac{ K (K-1)}{2 K - 1}
\end{align}
completing the converse for IC-CM.

\section{Achievability}
In this section, we provide achievability for the $K$-user IC-CM-EE (see Figure~\ref{fig:kic-subfigs}(c)), which will imply achievability for $K$-user IC-EE and $K$-user IC-CM. We will prove that,  for almost all channel gains, a sum secure d.o.f. lower bound of
\begin{equation}
D_{s,\Sigma} \ge \frac{ K (K-1)}{2 K - 1}
\label{ach-result}
\end{equation}
is achievable for the $K$-user IC-CM-EE.

\subsection{Background}
\label{sec:kic-achievable-scheme}

In this section, we will summarize the achievability scheme for the two-user IC-CM in \cite{xie_sdof_networks_in_prepare}, motivate the need for simultaneous alignment of multiple signals at multiple receivers in this $K$-user case, and provide an example of simultaneously aligning two signals at two receivers via asymptotic real alignment \cite{real_inter_align_exploit}. We provide the general achievable scheme for $K>2$ in Section~\ref{sec:kic-asymp-align} via cooperative jamming and asymptotic real alignment, and show that it achieves the sum secure d.o.f. in (\ref{ach-result})  via a detailed performance analysis in Section~\ref{sec:kic-performance-analysis}.

In the achievable scheme for $K=2$ in \cite{xie_sdof_networks_in_prepare}, four mutually independent discrete random variables $\{V_1,U_1,V_2,U_2\}$ are employed (see Figure~10 in \cite{xie_sdof_networks_in_prepare}). Each of them is uniformly and independently drawn from the discrete constellation $C(a,Q)$ given in (\ref{constel}). The role of $V_i$ is to carry message $W_i$, and the role of $U_i$ is to cooperatively jam receiver $i$ to help transmitter-receiver pair
$j$, where $j\neq i$, for $i,j=1,2$. By carefully selecting the
transmit coefficients, $U_1$ and $V_2$ are aligned at receiver $1$, and $U_2$ and $V_1$ are aligned at receiver $2$; and therefore, $U_1$
protects $V_2$, and $U_2$ protects $V_1$. By this signalling scheme,
information leakage rates are upper bounded by constants, and the
message rates are made to scale with power $P$, reaching the secure
d.o.f. capacity of the two-user IC-CM which is $\frac{2}{3}$.

\begin{figure}[t]
\centering
\includegraphics[scale=0.8]{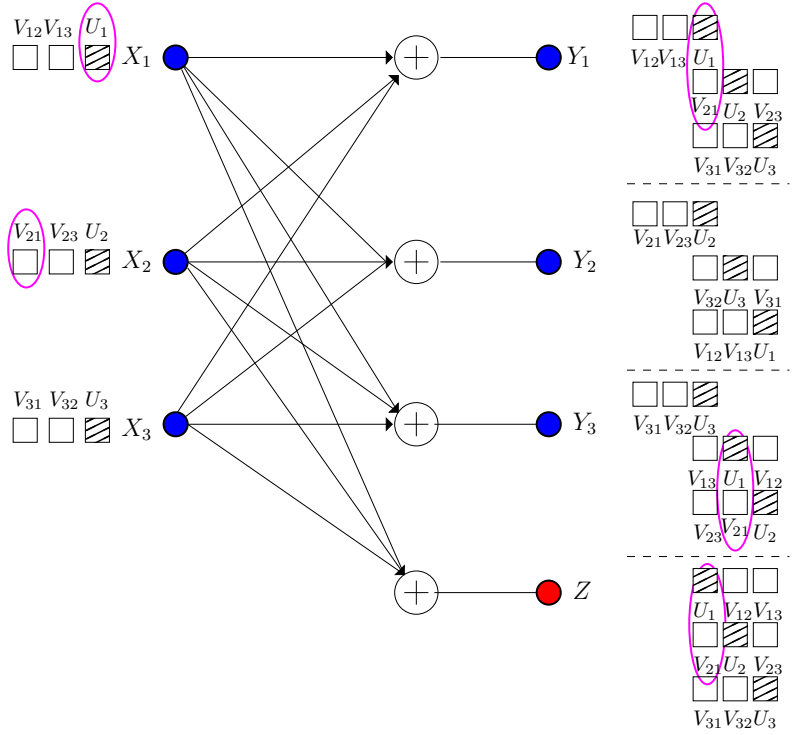}
\caption{Illustration of  alignment for $3$-user
IC-CM-EE. $U_1$ and $V_{21}$ are marked to emphasize their
simultaneous alignment at $Y_1$, $Y_3$ and $Z$. }
\label{fig:kic_alignment}
\end{figure}

Here, for the $K$-user IC-CM-EE, we employ a total of $K^2$ random variables,
\begin{align}
 V_{ij}, \quad & i,j = 1,\ldots,K, \, j\neq i \\
 U_k, \quad &k = 1,\ldots,K
\end{align}
which are illustrated in Figure~\ref{fig:kic_alignment} for the case of $K=3$. The scheme
proposed here has two major differences from
\cite{xie_sdof_networks_in_prepare}: 1) Instead of using a single
random variable to carry a message, we use a total of $K-1$ random
variables to carry each message. For transmitter $i$, $K-1$ random
variables $\{V_{ij}\}_{j\neq i}$, each representing a sub-message,
collectively carry message $W_i$. 2) Rather than protecting one
message at one receiver, each $U_k$ simultaneously protects a portion
of all sub-messages at all required receivers. More specifically,
$U_k$ protects  $\{V_{ik}\}_{i\neq k,i\neq j}$ at receivers $j$, and at the
eavesdropper (if there is any).  For example, in
Figure~\ref{fig:kic_alignment}, $U_1$ protects $V_{21}$ and $V_{31}$
where necessary. In particular, $U_1$ protects $V_{21}$ at receivers
1, 3 and the eavesdropper; and it protects $V_{31}$ at receivers 1, 2
and the eavesdropper. As a technical challenge, this requires $U_1$ to
be aligned with the same signal, say $V_{21}$, at multiple receivers
simultaneously, i.e., at receivers 1, 3 and the eavesdropper. These
particular alignments are circled by ellipsoids in Figure~\ref{fig:kic_alignment}. We do these simultaneous alignments using
asymptotic real alignment technique proposed in
\cite{real_inter_align_exploit} and used in
\cite{xie_k_user_ia_compound, interference_alignment_compound_channel}.

\begin{figure}[t]
\centering
\includegraphics[scale=0.76]{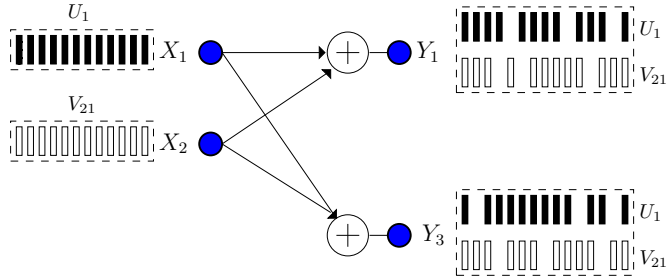}
\caption{Illustration of alignment at multiple receivers. }
\label{fig:kic_async}
\end{figure}

For illustration purposes, in the rest of this section, we demonstrate how we can align two signals at two receivers simultaneously; in particular, we will align $U_1$ with
$V_{21}$ at $Y_1$ and $Y_3$, simultaneously. Towards this
end, we will further divide the random variable $V_{21}$, which represents a sub-message,
into a large number of random variables denoted as $V_{21}\defn \{v_{21t}: t=1,\ldots,|T_{1}|\}$. We then send each one of these random variables after multiplying it with one of the coefficients in the following set which serves as the set of \emph{dimensions}:
\begin{align}
  T_{1}  = \Big\{
      h_{11}^{r_{11}}
      h_{21}^{r_{21}}
      h_{13}^{r_{13}}
      h_{23}^{r_{23}}
      :
      ~r_{11}, r_{21}, r_{13}, r_{23} \in \{1,\ldots,m\}
    \Big\}
\end{align}
where $m$ is a large constant. To perform the alignment, we let $U_1$ have the same
detailed structure as $V_{21}$, i.e., $U_1$ is also divided into a large
number of random variables as $U_{1}\defn \{u_{1t}:
t=1,\ldots,|T_{1}|\}$.  At receiver $1$, the elements of $U_1$ from
transmitter $1$ occupy the dimensions $h_{11} T_1$ and the elements of
$V_{21}$ from transmitter $2$ occupy the dimensions $h_{21} T_1$.
Although these two sets are not the same, their intersection contains
nearly as many elements as $T_1$, i.e.,
\begin{equation}
\label{eqn:kic-async-intersection}
\left| h_{11} T_1 \cap h_{21} T_1 \right| = m^2 (m-1)^2 \approx m^4 = |T_1|
\end{equation}
when $m$ is large, i.e., almost all elements of $U_1$ and $V_{21}$ are
asymptotically aligned at receiver $1$. The same argument applies for
receiver $3$. At receiver $3$, the elements of $U_1$ from
transmitter $1$ occupy the dimensions $h_{13} T_1$ and the elements of
$V_{21}$ from transmitter $2$ occupy the dimensions $h_{23} T_1$. Again,
although these two sets are not the same, their intersection contains
nearly as many elements as $T_1$. Therefore, almost all elements of $U_1$ and $V_{21}$ are aligned at receivers 1 and 3, simultaneously. These simultaneous alignments are depicted in Figure~\ref{fig:kic_async}. In the following section, we use this basic idea to align multiple signals at multiple receivers simultaneously. This will require a more intricate design of signals and dimensions.

\subsection{General Achievable Scheme via Asymptotic Alignment}
\label{sec:kic-asymp-align}

Here, we give the general achievable scheme for the $K$-user IC-CM-EE.
Let $m$ be a large constant. Let us define sets $T_i$, for $i=1,\ldots,K$, which will represent \emph{dimensions}
as follows:
\begin{align}
  T_{i} \defn \left\{
      h_{ii}^{r_{ii}}
      \left(
        \prod_{j,k=1, j\ne k}^{K} h_{jk}^{r_{jk}}
      \right)
      \left(
         \prod_{j=1}^{K} g_j^{s_{j}}
      \right)
      :
      ~r_{jk}, s_j \in \{1,\ldots,m\}
    \right\}
\end{align}
Let $M_i$ be the cardinality of $T_i$. Note that all $M_i$ are the same, thus we denote them as $M$,
\begin{equation}
  M \defn m^{1+K(K-1) + K} = m^{K^2 + 1}
\end{equation}
For each transmitter $i$, for $j\neq i$, let $\bft_{ij}$ be the vector containing all the elements in the set $T_j$. Therefore, $\bft_{ij}$ is an $M$-dimensional vector containing $M$ rationally independent real numbers in $T_j$. The sets $\bft_{ij}$ will represent the \emph{dimensions} along which message signals are transmitted. In particular, for any given $(i,j)$ with $i\neq j$, $\bft_{ij}$ will represent the dimensions in which message signal $V_{ij}$ is transmitted. In addition, for each transmitter $i$, let $\bft_{(i)}$ be the vector containing all the elements in the set $T_i$. Therefore, $\bft_{(i)}$ is an $M$-dimensional vector containing $M$ rationally independent real numbers in $T_i$. The sets $\bft_{(i)}$ will represent the \emph{dimensions} along which cooperative jamming signals are transmitted. In particular, for any given $i$, $\bft_{(i)}$ will represent the dimensions in which cooperative jamming signal $U_i$ is transmitted. Let us define a $KM$ dimensional vector $\mathbf{b}_i$ by stacking $\bft_{ij}$ and $\bft_{(i)}$ as
\begin{align}
\mathbf{b}_i^T
=
\left[
\mathbf{t}_{i1}^T,
\ldots,
\mathbf{t}_{i,i-1}^T,
\mathbf{t}_{i,i+1}^T,
\ldots,
\mathbf{t}_{iK}^T,
\mathbf{t}_{(i)}^T
\right]
\end{align}
Then, transmitter $i$ generates a vector $\mathbf{a}_i$, which
contains a total of $KM$ discrete signals each identically and independently
drawn from $C(a,Q)$. For convenience, we partition this transmitted signal as
\begin{align}
\mathbf{a}_i^T
=
\left[
\mathbf{v}_{i1}^T,
\ldots,
\mathbf{v}_{i,i-1}^T,
\mathbf{v}_{i,i+1}^T,
\ldots,
\mathbf{v}_{iK}^T,
\mathbf{u}_i^T
\right]
\end{align}
where $\bfv_{ij}$ represents the information symbols in $V_{ij}$, and $\bfu_i$
represents the cooperative jamming signal in $U_i$. Each of these vectors has length $M$, and therefore, the total length of $\mathbf{a}_i$ is $KM$. The channel input of transmitter $i$ is
\begin{equation}
x_i = \bfa_i^T \bfb_i
\end{equation}

Before we investigate the performance of this signalling scheme in
Section~\ref{sec:kic-performance-analysis}, we analyze the structure of
the received signal at the receivers. Without loss of generality we will focus on
receiver 1; by symmetry, a similar structure will exist
at all other receivers. We observe that in addition to the additive Gaussian noise,
receiver $1$ receives all the vectors $\bfv_{jk}$ for all $j,k\, (j\neq k)$ and
$\bfu_i$ for all $i$. All of these signals get multiplied with the corresponding channel gains
before they arrive at receiver 1. Due
to the specific signalling structure used at the transmitters, and
the multiplications with different channel gains over the wireless communication channel,
the signals arrive at the receiver lying in various different \emph{dimensions}.

To see the detailed structure of the received signals at the receivers, let us define $\tilde{T}_i$ as a superset of $T_i$, as follows
\begin{align}
  \tilde{T}_{i} \defn \left\{
      h_{ii}^{r_{ii}}
      \left(
         \prod_{j,k=1, j\ne k}^{K} h_{jk}^{r_{jk}}
      \right)
      \left(
         \prod_{j=1}^{K} g_j^{s_{j}}
      \right)
      :
      ~r_{jk}, s_j \in \{1,\ldots,m+1\}
    \right\}
\end{align}

The information symbols coming from transmitter 1 are in vectors $\bfv_{12}, \bfv_{13}, \ldots,  \bfv_{1K}$ which are multiplied by coefficients in $\bft_{12}, \bft_{13}, \ldots, \bft_{1K}$ before they are sent. These coefficients come from sets $T_2, T_3, \ldots, T_K$, respectively. After going through the channel, all of these coefficients get multiplied by $h_{11}$. Therefore, the receiving coefficients of $\bfv_{12}, \bfv_{13}, \ldots, \bfv_{1K}$ are $h_{11} \bft_{12}, h_{11} \bft_{13}, \ldots, h_{11} \bft_{1K}$, which are the \emph{dimensions} in the sets $h_{11} T_2, h_{11} T_3, \ldots, h_{11} T_K$, respectively. By construction, since each $T_i$ has powers of $h_{ii}$ in it (but no $h_{jj}$), these dimensions are {\it separate}. These correspond to {\it separate} boxes of $V_{12}$ and $V_{13}$ at receiver 1 in Figure~\ref{fig:kic_alignment} for the example case of $K=3$.

On the other hand, all of the cooperative jamming signals from all of the transmitters $\bfu_1, \bfu_2,\ldots,\bfu_K$ come to receiver 1 with received coefficients $h_{11} \bft_{(1)}, h_{21} \bft_{(2)},\ldots,h_{K1} \bft_{(K)}$, which are the \emph{dimensions} in the sets $h_{11} T_1, h_{21} T_2, \ldots, h_{K1} T_K$, respectively. We note that all of these dimensions are separate among themselves, and they are separate from the dimensions of the message signals coming from transmitter 1. That is, all of the dimensions in $h_{11} T_2, h_{11} T_3, \ldots, h_{11} T_K$ and $h_{11} T_1, h_{21} T_2, \ldots, h_{K1} T_K$ are all mutually different, again owing to the fact that each $T_i$ contains powers of $h_{ii}$ in it. These correspond to separate boxes of $V_{12}$, $V_{13}$, $U_1$, $U_2$ and $U_3$ at receiver 1 in Figure~\ref{fig:kic_alignment} for the example case of $K=3$.

Next, we note that each $\bfu_i$ is aligned together with all of the $\bfv_{ji}$ coming from the $j$th transmitter, with $j \neq i$ and $j \neq 1$, at receiver 1. Note that $\bfu_i$ occupies dimensions $h_{i1} T_i$ and $\bfv_{ji}$ (for any $j \neq i$ and $j \neq 1$) occupies dimensions $h_{j1} T_i$ at receiver 1. From the arguments in Section~\ref{sec:kic-achievable-scheme}, $\bfu_i$ and $\bfv_{ji}$ (with $j\neq i$ and $j\neq 1$) are asymptotically aligned. More formally, we note that $\bfu_i$ occupies dimensions $h_{i1} T_i$ which is contained in $\tilde{T}_i$. Similarly, all $\bfv_{ji}$, with $j \neq i$ and $j\neq 1$, occupy dimensions $h_{j1} T_i$, respectively, which are all contained in $\tilde{T}_i$. Therefore, $\bfu_i$ and all $\bfv_{ji}$ (with $j \neq i$ and $j \neq 1$) are all aligned along $\tilde{T}_i$. These alignments are shown as $U_1$ being aligned with $V_{21}$ and $V_{31}$; $U_2$ being aligned with $V_{32}$; and $U_3$ being aligned with $V_{23}$ at receiver 1 in Figure~\ref{fig:kic_alignment} for the example case of $K=3$. Further, we note that, since only $T_i$ and $\tilde{T}_i$ contain powers of $h_{ii}$, the dimensions $h_{11} T_2, h_{11} T_{3}, \ldots, h_{11} T_K, \tilde{T}_1, \tilde{T}_2,\ldots,\tilde{T}_K$ are all separable. This implies that all the elements in the set
\begin{equation}
  R_1 \defn
  \left(
    \displaystyle \bigcup_{j=2}^K h_{11} T_j
  \right)
  \bigcup
  \left(
    \displaystyle \bigcup_{j=2}^K \tilde{T}_j
  \right)
  \bigcup
    \tilde{T}_1
\end{equation}
are rationally independent, and thereby the cardinality of $R_1$ is
\begin{align}
M_R
\defn |R_1|
& = (K-1) m^{1 + K(K-1) + K} + K (m+1)^{1 + K(K-1) + K} \\
& = (K-1) m^{K^2+1} + K (m+1)^{K^2+1}
\end{align}

\subsection{Performance Analysis}
\label{sec:kic-performance-analysis}
We will compute the secrecy rates achievable with the asymptotic alignment based scheme proposed in Section~\ref{sec:kic-asymp-align} by using the following theorem.
\begin{theorem}
\label{kic:theo-achievability}
For $K$-user interference channels with confidential messages and one
external eavesdropper, the following rate region is achievable
\begin{equation}
R_i\ge I(V_i;Y_i) - {\max_{j\in\mathcal{K}_{0,-i}}} I(V_i; Y_j|V_{-i}^K), \qquad i=1,\ldots,K
\label{eqn:kic-lower-bound}
\end{equation}
where for convenience we denote $Z$ by $Y_0$, $V_{-i}^K \defn \{V_j\}_{j=1,j\neq i}^K$ and $\mathcal{K}_{0,-i} = \{0,1,\ldots, i-1,i+1, \ldots,K\}$. The auxiliary random variables
$\{V_i\}^K_{i=1}$ are mutually independent, and for each $i$, we
have the following Markov chain $V_i\rightarrow X_i\rightarrow
(Y_1,\ldots,Y_K)$.
\end{theorem}

In developing the achievable rates in Theorem~\ref{kic:theo-achievability}, we focus on a single transmitter, say $i$, and consider the compound setting associated with message $W_i$, where this message needs to be secured against a total of $K$ eavesdroppers, with $K-1$ of them being the other legitimate receivers ($j\neq i$) and the remaining one being the external eavesdropper ($j=0$). A proof of this theorem is given in Appendix~\ref{kic:appendix-proof-achievability}.

We apply Theorem~\ref{kic:theo-achievability} to our alignment based scheme proposed in Section~\ref{sec:kic-asymp-align} by selecting $V_i$ used in (\ref{eqn:kic-lower-bound}) as
\begin{align}
V_i \defn (\bfv_{i1}^T, \ldots, \bfv_{i,i-1}^T, \bfv_{i,i+1}^T ,\ldots, \bfv_{iK}^T)
\end{align}
for $i=1,\ldots,K$.  For any $\delta>0$, if we choose $Q =
 P^{\frac{1-\delta}{2(M_R+\delta)}}$ and $a = \frac{\gamma
P^{\frac{1}{2}}}{Q}$,  based on Lemma~\ref{lemma:ria_real_alignment},
the probability of error of estimating $V_i$ based on $Y_i$
can be upper bounded by a function decreasing
exponentially fast in $P$, by choosing a $\gamma$, a positive constant
independent of $P$ to normalize the average power of the input
signals, as
\begin{align}
 0 < \gamma \le \frac{1}{ {\sum_{t\in\mathbf{b}_i} |t| }}
  = \frac{1}{ {\sum_{i=1}^{K} \sum_{t_i\in T_i} |t_i|} }
\end{align}
Furthermore, by Fano's inequality, we can conclude that
\begin{align}
  I(V_i; Y_i) & \ge
\frac{ (K-1) m^{K^2+1} (1-\delta)} { M_R + \delta }
\left(\frac{1}{2} \log P\right) + o(\log P)\\
& = \frac{ (K-1)   (1-\delta)} { K -1  + K \left( 1 + \frac{1}{m}
\right)^{K^2+1} + \frac{\delta}{m^{K^2+1}} }
\left(\frac{1}{2} \log P\right) + o(\log P)
\label{kic-ivy_low}
\end{align}
where $o(\cdot)$ is the little-$o$ function. This provides a lower bound for the first term in \eqn{eqn:kic-lower-bound}.

Next, we need to derive an upper bound for the second item in \eqn{eqn:kic-lower-bound}, i.e, the secrecy penalty. For any $i\in\mk=\{1,\ldots,K\}$ and $j\in\mki=\{1,\ldots,i-1,i+1,\ldots,K\}$, by the Markov chain $V_i \rightarrow (\sum_{k=1}^K h_{kj}X_{kj} , V_{-i}^K) \rightarrow Y_j$,
\begin{align}
 I(V_i; Y_j | V_{-i}^K )
 & \le I\left( V_i; \sum_{k=1}^K h_{kj} X_k \Big| V_{-i}^K \right) \\
 & =  H\left( \sum_{k=1}^K h_{kj} X_k \Big| V_{-i}^K \right)
     - H\left(\sum_{k=1}^K h_{kj} X_k \Big| V_1^K \right)
 \label{eqn:kic-secrecy-vi-yj}
\end{align}
where $V_1^K=\{V_1,\ldots,V_K\}$. The first term in \eqn{eqn:kic-secrecy-vi-yj} can be rewritten as
\begin{align}
H\left( \sum_{k=1}^K h_{kj} X_k \Big| V_{-i}^K \right)
& = H\left({
  \sum_{k=1}^K h_{kj} \bfu_k^T \bft_{(k)} +
  \mathop{
    \sum_{k=1}^K
  }_{k\neq i} h_{ij}\bfv_{ik}^T \bft_{ik}
} \right) \\
& = H\left({
   h_{ij} \bfu_i^T \bft_{(i)} +
  \mathop{
    \sum_{k=1}^K
  }_{k\neq i}
  \Big[
    h_{ij}\bfv_{ik}^T \bft_{ik} + h_{kj} \bfu_k^T \bft_{(k)}
  \Big]
} \right)
\end{align}
Note that, for a given $k$, the vectors $\bft_{ik}$ and $\bft_{(k)}$
represent the same \emph{dimensions} $T_k$, and $h_{ij}, h_{kj} \in
T_k$ for all $k\neq i$, which implies that $h_{ij} T_k, h_{kj} T_k
\in \tilde{T}_k$. In addition, for each $k$, we note that a large part of
the two sets $h_{ij} T_k$ and $h_{kj} T_k$ are the same, i.e.,
\begin{equation}
 \left|
   h_{ij} T_k \bigcap h_{kj} T_k
   \right| = m^{K^2-1} (m-1)^2 \defn M_{\delta}
 \label{eqn:kic-cadinarlity-intersection}
\end{equation}
Therefore, the first term in \eqn{eqn:kic-secrecy-vi-yj} can be further upper bounded as
\begin{align}
H\left( \sum_{k=1}^K h_{kj} X_k \Big| V_{-i}^K \right)
& = H\left({
   h_{ij} \bfu_i^T \bft_{(i)} +
  \mathop{
    \sum_{k=1}^K
  }_{k\neq i}
  \Big[
    h_{ij}\bfv_{ik}^T \bft_{ik} + h_{kj} \bfu_k^T \bft_{(k)}
  \Big]
} \right) \\
& \le
\log \Big[ (2Q+1)^M (4Q+1)^{(K-1) M_\delta} (2Q+1)^{2(K-1)(M-M_\delta)}
\Big] \\
& \le
\log \Big[ Q^{M + (K-1) M_\delta + 2(K-1)(M-M_\delta)}
\Big] + o(\log P) \\
& \le
 \frac{
\left[M + (K-1) M_\delta + 2(K-1)(M-M_\delta)\right] (1-\delta)
}{
(K-1) m^{K^2+1} + K (m+1)^{K^2+1} + \delta
} \left( \frac{1}{2} \log P \right) \nl
& \quad + o(\log P) \\
& \le
 \frac{
  \left\{ 1 + (K-1) \left( 1 - \frac{1}{m}\right)^2 + 2(K-1)\left[1-\left(
    1 - \frac{1}{m}\right)^2\right] \right\}  (1-\delta)
}{
K -1  + K \left( 1 + \frac{1}{m} \right)^{
K^2+1} + \frac{\delta}{m^{K^2+1}}
} \left( \frac{1}{2} \log P \right) \nl
& \quad + o(\log P)
 \label{eqn:kic-secrecy-vi-yj-1}
\end{align}

The second term in \eqn{eqn:kic-secrecy-vi-yj} is exactly the
entropy of $\{\bfu_k\}_{k=1}^K$ vectors, i.e.,
\begin{align}
H\left(\sum_{k=1}^K h_{kj} X_k | V_1^K \right)
& = H\left( \sum_{k=1}^K h_{kj} \bfu_{k}^T \bft_{(k)} \right)  \\
& = \log (2Q+1)^{K M} \\
& = \frac{K m^{K^2+1} (1-\delta)}{(K-1) m^{K^2+1} + K (m+1)^{K^2+1} + \delta
} \left( \frac{1}{2} \log P\right)\nl
& \quad + o(\log P)\\
& =
 \frac{
 K (1-\delta)
}{
K -1  + K \left( 1 + \frac{1}{m} \right)^{
K^2+1} + \frac{\delta}{m^{K^2+1}}
} \left( \frac{1}{2} \log P\right) + o(\log P)
 \label{eqn:kic-secrecy-vi-yj-2}
\end{align}
Substituting \eqn{eqn:kic-secrecy-vi-yj-1} and
\eqn{eqn:kic-secrecy-vi-yj-2} into \eqn{eqn:kic-secrecy-vi-yj}, we get
\begin{align}
 I(V_i; Y_j | V_{-i}^K )
 & \le  H\left( \sum_{k=1}^K h_{kj} X_k \Big| V_{-i}^K \right)
     - H\left(\sum_{k=1}^K h_{kj} X_k \Big| V_1^K \right) \\
 & \le
 \frac{
   \left\{ 1 + (K-1) \left( 1 - \frac{1}{m}\right)^2 + 2(K-1)\left[1-\left(
   1 - \frac{1}{m}\right)^2\right] - K \right\}  (1-\delta)
}{
K -1  + K \left( 1 + \frac{1}{m} \right)^{
K^2+1} + \frac{\delta}{m^{K^2+1}}
} \left( \frac{1}{2} \log P \right) \nl
& \quad + o(\log P) \\
 & \le
 \frac{
   K \,\frac{2m-1}{m^2} (1-\delta)
}{
K -1  + K \left( 1 + \frac{1}{m} \right)^{
K^2+1} + \frac{\delta}{m^{K^2+1}}
} \left( \frac{1}{2} \log P \right)  + o(\log P)
 \label{eqn:kic-secrecy-vi-yj-final}
\end{align}
We note that by choosing $m$ large enough, the factor before the
$\frac{1}{2}\log P$ term can be made arbitrarily small. Due to the non-perfect (i.e., only asymptotical) alignment, the upper bound for the information leakage rate is not a constant as in \cite{xie_sdof_networks_in_prepare}, but a function which can be made to approach zero d.o.f.

For any $i\in\mk$ and $j=0$, i.e., $Y_0=Z$ the external eavesdropper,
we should derive a new upper bound for the second term in \eqn{eqn:kic-secrecy-vi-yj}, i.e., $I(V_i;Z|V_{-i}^K)$. By similar steps, we have
\begin{align}
I(V_i;Z|V_{-i}^K)
& \le
I\left( V_i; \sum_{k=1}^K g_k X_k  \Big|V_{-i}^K \right) \\
& =
H\left(\sum_{k=1}^K g_k X_k  \Big|V_{-i}^K \right)
-
H\left(\sum_{k=1}^K g_k X_k
\Big|V_{1}^K\right)  \\
& =
H\left(\sum_{k=1}^K g_k X_k  \Big|V_{-i}^K \right)
-
 H\left( \sum_{k=1}^K g_{k} \bfu_{k}^T \bft_{(k)} \right)  \\
& =
H\left(\sum_{k=1}^K g_k X_k  \Big|V_{-i}^K \right)
-
 \log (2Q+1)^{K M}
 \label{eqn:kic-secrecy-vi-z}
\end{align}
Here, we need to upper bound the first item in \eqn{eqn:kic-secrecy-vi-z}.
We first observe that
\begin{align}
H\left(\sum_{k=1}^K g_k X_k  \Big|V_{-i}^K \right)
& = H\left(
\sum_{k=1}^K g_{k} \bfu_{k}^T \bft_{(k)}
+
\mathop{ \sum_{k=1}^K }_{k\neq i} g_i \bfv_{ik}^T \bft_{ik}
\right) \\
& = H\left(
 g_{i} \bfu_{i}^T \bft_{(i)}
+
\mathop{ \sum_{k=1}^K }_{k\neq i}
\Big[
  g_{k} \bfu_{k}^T \bft_{(k)} + g_i \bfv_{ik}^T \bft_{ik}
\Big]
\right)
\end{align}
Firstly, note that, $\bft_{(k)}$ and $\bft_{ik}$ represent the same set
$T_k$. Therefore, for different $k$, the
\emph{dimensions} are distinguishable. Secondly, due to reasons similar
to \eqn{eqn:kic-cadinarlity-intersection}, we conclude that
\begin{align}
H\left(\sum_{k=1}^K g_k X_k  \Big|V_{-i}^K \right)
& = H\left(
 g_{i} \bfu_{i}^T \bft_{(i)}
+
\mathop{ \sum_{k=1}^K }_{k\neq i}
\Big[
  g_{k} \bfu_{k}^T \bft_{(k)} + g_i \bfv_{ik}^T \bft_{ik}
\Big]
\right) \\
& \le
\log \Big[ (2Q+1)^M (4Q+1)^{(K-1) M_\delta} (2Q+1)^{2(K-1)(M-M_\delta)}
\Big] \\
& \le
\log \Big[ Q^{M + (K-1) M_\delta + 2(K-1)(M-M_\delta)}
\Big] + o(\log P) \\
& \le
 \frac{
\left[M + (K-1) M_\delta + 2(K-1)(M-M_\delta)\right] (1-\delta)
}{
(K-1) m^{K^2+1} + K (m+1)^{K^2+1} + \delta
} \left( \frac{1}{2} \log P \right) \nl
& \quad + o(\log P)
 \label{eqn:kic-secrecy-vi-z-1}
\end{align}
Substituting \eqn{eqn:kic-secrecy-vi-z-1} into
\eqn{eqn:kic-secrecy-vi-z}, we attain an upper bound which is the same as the
upper bound for $I(V_i;Y_j|V_{-i}^K)$, i.e.,
\begin{align}
I(V_i;Z|V_{-i}^K)
 & \le
 \frac{
   K \,\frac{2m-1}{m^2} (1-\delta)
}{
K -1  + K \left( 1 + \frac{1}{m} \right)^{
K^2+1} + \frac{\delta}{m^{K^2+1}}
} \left( \frac{1}{2} \log P \right)  + o(\log P)
 \label{eqn:kic-secrecy-vi-z-final}
\end{align}

Substituting \eqn{kic-ivy_low}, \eqn{eqn:kic-secrecy-vi-yj-final}, and
\eqn{eqn:kic-secrecy-vi-z-final} into
\eqn{eqn:kic-lower-bound}, we obtain a lower bound for the achievable
secrecy rate $R_i$ as
\begin{equation}
  R_i \ge \frac{\left[ (K-1) - K \left( \frac{2m-1}{m^2}\right)\right] (1-\delta)
  } { K -1  + K \left( 1 + \frac{1}{m}
  \right)^{K^2+1} + \frac{\delta}{m^{K^2+1}} }
\left(\frac{1}{2} \log P\right) + o(\log P)
\end{equation}
By choosing  $m\to\infty$ and $\delta\to0$, we can achieve secrecy sum rates
arbitrarily close to $\frac{K-1}{2K-1}\left( \frac{1}{2} \log
P\right)$, thereby achieving
the sum secure d.o.f. lower bound in (\ref{ach-result}).

\section{Conclusion}
In this paper, we studied secure communications in $K$-user Gaussian interference networks from an information-theoretic point of view, and addressed three important channel models: IC-EE, IC-CM and their combination IC-CM-EE in a unified framework. We showed that, for all three models, the sum secure d.o.f.
is exactly $\frac{K(K-1)}{2K-1}$. Our achievability is based on structured signalling,
structured cooperative jamming, channel prefixing and asymptotic real interference alignment.
The key insight of the achievability is to carefully design the structure of all of the signals at the transmitters so that the signals are received at both legitimate receivers and eavesdroppers in most desirable manner from a secure communication point of view. In particular, cooperative jamming signals protect information carrying signals via alignment, and the information carrying signals are further aligned to maximize secure d.o.f.

\appendix

\section{Proof of Theorem \ref{kic:theo-achievability}}
\label{kic:appendix-proof-achievability}
We first provide an outline of the proof. Our proof will combine and extend techniques from \cite{secrecy_ic} and \cite{compound_wiretap_channel}. Our approach has three main components.
First, as in \cite{secrecy_ic}, we condition the mutual information representing the secrecy leakage rate on the signals that carry the messages of other transmitter-receiver pairs. That is, for any given $i$, we condition the subtracted mutual information term in (\ref{eqn:kic-lower-bound}) on $V_{-i}^K$. This creates {\it enhanced} eavesdroppers. If we can guarantee secrecy against these enhanced eavesdroppers, we can guarantee secrecy against the original eavesdroppers. More specifically, for the leakage rate of message of transmitter $i$ at receiver $j$, with $j\neq i$, we use
\begin{equation}
I(V_i; Y_j|V_{-i}^K)  = I(V_i; Y_j,V_{-i}^K) \defn I(V_i; \tilde Y_j )
\end{equation}
where $\tilde Y_j \defn (Y_j,V_{-i}^K)$ is the output of an {\it enhanced} eavesdropper with respect to message $W_i$. Second, as in \cite{compound_wiretap_channel}, we consider the secrecy rate achievable against the {\it strongest} enhanced eavesdropper for each message. Therefore, as argued in \cite[Appendex A]{compound_wiretap_channel}, if we can guarantee a secrecy rate against the strongest eavesdropper, we can guarantee this secrecy rate against the original eavesdroppers. More specifically, let $Y^{(i)}$ be an element of the set $\{Y_1,\ldots,Y_k,Z\} \backslash \{Y_i\}$ such that
\begin{equation}
\label{eqn:kic-secrecy-rate-lb}
  I(V_i;Y^{(i)}|V_{-i}^K) = \max_{j\in\mkzi} (V_i;Y_j|V_{-i}^K)
\end{equation}
That is, $Y^{(i)}$ is the \emph{strongest} eavesdropper with respect to transmitter $i$. The achievable rate in (\ref{eqn:kic-lower-bound}) considers the strongest eavesdropper for each message. Therefore, for each transmitter $i$, we construct a compound wiretap code as in \cite{compound_wiretap_channel}. Third, we prove secrecy for each message $W_i$, via the following equivocation inequality
\begin{align}
\label{eqn:kic-secrecy-compound}
  H(W_i|\bfY^{(i)},\bfvmi^K)  \ge H(W_i) - n \epsilon^{(i)}, \qquad i=1,\ldots,K
\end{align}
for some arbitrarily small number $\epsilon^{(i)}$. Here, as in the main body of the paper, we denote $n$-length sequences with boldface letters. The secrecy constraints in (\ref{eqn:kic-secrecy-compound}) fit the created equivalent view of the channel better. As we show next, secrecy constraints in (\ref{eqn:kic-secrecy-compound}) imply our original secrecy constraints in \eqn{eqn:kic-secrecy-constraint-cmee-1} and \eqn{eqn:kic-secrecy-constraint-cmee-2}.

Towards this end, first note that, for each $i$,
\begin{align}
 H(W_i|\bfY_j,\bfvmi^K) \ge H(W_i|\bfY^{(i)},\bfvmi^K)  \ge H(W_i) - n \epsilon^{(i)}
\end{align}
for all $j\in\mkzi$ since $Y^{(i)}$ is the \emph{strongest} eavesdropper with respect to transmitter $i$ and by using the enhanced eavesdropper argument in \cite[Appendex A]{compound_wiretap_channel}. Then, the fact that \eqn{eqn:kic-secrecy-compound} for all $i$ implies the original secrecy constraints in \eqn{eqn:kic-secrecy-constraint-cmee-1} and
\eqn{eqn:kic-secrecy-constraint-cmee-2} follows from the following derivation:
\begin{align}
  H(W_{-j}^K|\bfY_j)
  & \ge  H(W_{-j}^K|\bfY_j, W_j) \\
  & \ge  \sum_{i\neq j}H(W_{i}|\bfY_j, W_{-i}^K) \\
  & \ge  \sum_{i\neq j}H(W_{i}|\bfY_j, \bfV_{-i}^K, W_{-i}^K) \\
  & = \sum_{i\neq j}H(W_{i}|\bfY_j, \bfV_{-i}^K)
  \label{eqn:kic-secrecy-cal-markov} \\
  & \ge \sum_{i\neq j}H(W_{i}|\bfY^{(i)}, \bfV_{-i}^K) \\
  & \ge \sum_{i\neq j} \Big[ H(W_{i}) - n \epsilon^{(i)} \Big] \\
  & =  H(W_{-j}^K) - n \epsilon^{(-j)}
\end{align}
where \eqn{eqn:kic-secrecy-cal-markov} is due to the Markov chain
$W_{-i}^K \rightarrow (\bfY_j, \bfV_{-i}^K) \rightarrow W_i$. Similarly,
\begin{align}
  H(W^K|\bfZ)
  & \ge  \sum_{i}H(W_{i}|\bfZ, W_{-i}^K) \\
  & \ge  \sum_{i}H(W_{i}|\bfZ, \bfV_{-i}^K, W_{-i}^K) \\
  & = \sum_{i}H(W_{i}|\bfZ, \bfV_{-i}^K) \\
  & \ge \sum_{i}H(W_{i}|\bfY^{(i)}, \bfV_{-i}^K) \\
  & \ge \sum_{i} \Big[ H(W_{i}) - n \epsilon^{(i)} \Big] \\
  & =  H(W^K) - n \epsilon^{(Z)}
\end{align}
where $\epsilon^{(Z)}$ is small for sufficiently large $n$.

We start by choosing the following rates for the secure and confusion messages of transmitter $i$:
\begin{align}
  R_i  & = I(V_i;Y_i) - I(V_i;Y^{(i)}|V_{-i}^K) - \epsilon \\
  R_i^c  & = I(V_i;Y^{(i)}|V_{-i}^K) - \epsilon
\end{align}
Transmitter $i$ generates $2^{n(R_i+R_i^c)}$ independent sequences
each with probability
\begin{equation}
  p(\mathbf{v}_i) = \prod_{t=1}^n p(v_{it})
\end{equation}
and constructs a codebook as
\begin{equation}
  C_i \defn \Big\{
    \mathbf{v}_i(w_i,w_i^c) \,:\,
    w_i\in \{1,\ldots,2^{n R_i}\},
    w_i^c\in \{1,\ldots,2^{n R_i^c}\}
  \Big\}
\end{equation}
To transmit a message $w_i$, transmitter $i$ chooses an element $\mathbf{v}_i$ from the sub-codebook $C_i(w_i)$
\begin{equation}
  C_i(w_i) \defn \Big\{
    \mathbf{v}_i(w_i,w_i^c) \,:\,
    w_i^c\in \{1,\ldots,2^{n R_i^c}\}
  \Big\}
\end{equation}
and generates a channel input sequence based on
\begin{equation}
  p(x_i|v_i)
\end{equation}

Due to the code construction, we have $R_i + R_i^c < I(V_i;Y_i)$, for all
$i$. Therefore, for sufficiently large $n_i$, we can find a
codebook such that the probability of error at the corresponding receiver $i$  can be upper
bounded by an arbitrarily
small number, i.e., $\pr(e_i)^{(n_i)} \le \epsilon$. Then, let
$n=\max_{i} n_i$, which gives
$\max_{i}\pr(e_i)^{(n)}\le\epsilon$.

For the equivocation calculation, we consider the following conditional entropy as discussed before:
\begin{align}
  H(W_i|\bfY^{(i)},\bfvmi^K)
  = & H(W_i, \bfY^{(i)}| \bfvmi^K) - H(\bfY^{(i)}|\bfvmi^K) \label{eqn:kic-equivocation-start}\\
  = &  H(W_i, \bfvi, \bfY^{(i)}|\bfvmi^K)
       - H(\bfvi| W_i, \bfY^{(i)}, \bfvmi^K)
       - H(\bfY^{(i)}|\bfvmi^K)\\
  = & H(W_i, \bfvi|\bfvmi^K) + H(\bfY^{(i)} | W_i, \bfvi, \bfvmi^K)
       - H(\bfvi| W_i, \bfY^{(i)}, \bfvmi^K) \nonumber \\
    &  - H(\bfY^{(i)}|\bfvmi^K)\\
  = & H(W_i, \bfvi|\bfvmi^K)
       - H(\bfvi| W_i, \bfY^{(i)}, \bfvmi^K)
       + H(\bfY^{(i)} | \bfvi, \bfvmi^K)
       - H(\bfY^{(i)}|\bfvmi^K)
  \label{eqn:kic-equivocation}
\end{align}
where the last equality is due to the Markov chain $W_i \rightarrow
(\bfvi, \bfvmi^K) \rightarrow \bfY^{(i)} $.

The first term in \eqn{eqn:kic-equivocation} is exactly the entropy of
codebook $C_i$
\begin{equation}
  H(\bfvi) = n (R_i + R_i^c)
  \label{eqn:kic-equivocation-1}
\end{equation}

To bound the second term in \eqn{eqn:kic-equivocation}, we have the
following observation: Given the message $W_i = w_i$ and the received
sequences $\bfY^{(i)}=\mathbf{y}^{(i)}$ and genie-aided sequences
$\bfvmi^K=\mathbf{v}_{-i}^K$, receiver $Y^{(i)}$ can decode the codeword
$\mathbf{v}_i(w_i,w_i^c)$ with arbitrarily small probability of error
$\lambda(w_i)^{(n)}$ as $n$ gets very large. More formally: by giving $W_i = w_i, \bfvmi^K=\mathbf{v}_{-i}^K$, receiver
$Y^{(i)}$ decodes $\bfvi$ if there is a unique $w_i^c$ such that
\begin{equation}
  \big(\mathbf{v}_i(w_i,w_i^c), \mathbf{y}^{(i)}\big) \in
  T_{\epsilon}^{(n)}(P_{V_1,Y^{(i)}|V_{-i}^K})
\end{equation}
Otherwise, the receiver declares an error. Without loss of generality, we assume that
$\mathbf{v}_i(w_i,w_1^c)$ is sent and denote the event $
\left\{ (\mathbf{v}_i(w_i,w_j^c), \mathbf{y}^{(i)}) \in
T_{\epsilon}^{(n)}(P_{V_1,Y^{(i)}|V_{-i}^K})\right\} $ as $E_{j}$.
Therefore, the probability of error $\lambda(w_i)^{(n)}$ can be
bounded as
\begin{align}
\lambda(w_i)^{(n)}
\le
\pr(E_1^c) + \sum_{j\neq 1} \pr(E_j)
\end{align}
where the probability here is
conditioned on the event that $\mathbf{v}_i(w_i,w_1^c)$ is sent. By joint
typicality, we know that $\pr(E_1^c) \le \nextep$ for sufficiently
large $n$, and
\begin{equation}
  \pr(E_j)
  \le 2^{n H(V_i,Y^{(i)}|V_{-i}^K)-n H(V_i) - nH(Y^{(i)}|V_{-i}^K) - n\nextep }
  = 2^{- n I(V_I;Y^{(i)}|V_{-i}^K) -n \nextepnu }
\end{equation}
Hence,
\begin{align}
\lambda(w_i)^{(n)}
\le  \nexteppre + 2^{nR_i^c} 2^{- n I(V_I;Y^{(i)}|V_{-i}^K) - n\nextepnu }
\end{align}
Note that $R_i^c  = I(V_i;Y^{(i)}|V_{-i}^K) - \epsilon$. Therefore, we can
conclude that $\lambda(w_i)^{(n)}\le \nextep $ for sufficiently large
$n$, which by Fano's inequality further implies that
\begin{align}
H(\bfvi| W_i, \bfY^{(i)}, \bfvmi^K)
= \sum_{W_i=w_i, \bfY^{(i)} = \mathbf{y}^{(i)}, \bfvmi^K = \mathbf{v}^K_{-i}}
H(\bfvi| w_i, \mathbf{y}^{(i)}, \mathbf{v}_{-i}^K)
\le n \nextep
\label{eqn:kic-equivocation-2}
\end{align}

The third term in \eqn{eqn:kic-equivocation} can be lower bounded as follows:
\begin{align}
 H(\bfY^{(i)}|\bfvi,\bfvmi^K)
 & = \sum_{\mathbf{v}_i,\mathbf{v}_{-i}^K}
 \pr(\bfvi=\mathbf{v}_i )\pr(\bfvmi^K=\mathbf{v}_{-i}^K)
 H(\bfY^{(i)}|\bfvi=\mathbf{v}_i,\bfvmi^K=\mathbf{v}_{-i}^K) \\
 & \ge \sum_{(\mathbf{v}_i,\mathbf{v}_{-i}^K)\in
 T^{(n)}_{\epsilon}(P_{V_i,V_{-i}^K})}
 \Bigg[ \pr(\bfvi=\mathbf{v}_i )\pr(\bfvmi^K=\mathbf{v}_{-i}^K) \nonumber\\
 & \quad\quad \quad\quad\quad\quad\quad\quad\quad\quad\quad\quad
 H(\bfY^{(i)}|\bfvi=\mathbf{v}_i,\bfvmi^K=\mathbf{v}_{-i}^K) \Bigg]\\
 & \ge \sum_{(\mathbf{v}_i,\mathbf{v}^K_{-i})\in
 T^{(n)}_{\epsilon}(P_{V_i,V_{-i}^K})}
 \Bigg[ \pr(\bfvi=\mathbf{v}_i )\pr(\bfvmi^K=\mathbf{v}_{-i}^K) \nl
 & \quad\quad\quad\quad\quad
   \sum_{(a,b)\in\mathcal{V}_i\times\mathcal{V}_{-i}^K}
   N(a,b|\mathbf{v}_{i},\mathbf{v}_{-i}^K)
   \sum_{y^{(i)}\in\mathcal{Y}^{(i)}} - p(y^{(i)}|a,b) \log p(y^{(i)}|a,b)
 \Bigg]
 \\
 & \ge \sum_{(\mathbf{v}_i,\mathbf{v}_{-i}^K)\in
 T^{(n)}_{\epsilon}(P_{V_i,V_{-i}^K})}
 \Bigg[ \pr(\bfvi=\mathbf{v}_i )\pr(\bfvmi^K=\mathbf{v}^K_{-i}) \nl
 & \quad\quad\quad\quad\quad\quad
   \sum_{(a,b)\in\mathcal{V}_i\times\mathcal{V}_{-i}^K}
   n \Big(\pr(V_i =a ,V_{-i}^K=b) - \nextep\Big) \nonumber \\
   & \quad\quad\quad\quad\quad\quad\quad\quad\quad
   \sum_{y^{(i)}\in\mathcal{Y}^{(i)}} - p(y^{(i)}|a,b) \log p(y^{(i)}|a,b)
 \Bigg]
 \\
 & \ge \sum_{(\mathbf{v}_i,\mathbf{v}_{-i}^K)\in
 T^{(n)}_{\epsilon}(P_{V_i,V_{-i}^K})}
 n\Big[ \pr(\bfvi=\mathbf{v}_i )\pr(\bfvmi^K=\mathbf{v}^K_{-i})
    H(Y^{(i)}|V_i, V_{-i}^K) - \nextep
 \Big] \\
 & \ge (1-\nextep) n H(Y^{(i)}|V_i, V_{-i}^K) -n \nextep \\
 & \ge  n H(Y^{(i)}|V_i, V_{-i}^K) - n \nextep
\label{eqn:kic-equivocation-3}
\end{align}

To compute the forth term in \eqn{eqn:kic-equivocation}, we define
\begin{equation}
  \hat{\bfY}^{(i)}  = \left\{
    \begin{array}{ll}
      \bfY^{(i)}, &  \textrm{if } (\mathbf{v}_{-i}^K,\mathbf{y}^{(i)})
      \in T_{\epsilon}^{(n)} (P_{V_{-i}^K,Y^{(i)}}) \\
      \textrm{arbitrary}, & \textrm{otherwise}
    \end{array}
  \right.
\end{equation}
Then, we obtain
\begin{align}
  H(\bfY^{(i)}|\bfV_{-i}^K)
  & = \sum_{\mathbf{v}_{-i}^K} \pr(\bfV_{-i}^K=\mathbf{v}_{-i}^K)
    H(\bfY^{(i)}| \bfV_{-i}^K=\mathbf{v}_{-i}^K) \\
  & \le \sum_{\mathbf{v}_{-i}^K} \pr(\bfV_{-i}^K=\mathbf{v}_{-i}^K)
    H(\bfY^{(i)},\hat{\bfY}^{(i)}| \bfV_{-i}^K=\mathbf{v}_{-i}^K) \\
  & = \sum_{\mathbf{v}_{-i}^K} \pr(\bfV_{-i}^K=\mathbf{v}_{-i}^K)
  \Big[
    H(\hat{\bfY}^{(i)}| \bfV_{-i}^K=\mathbf{v}_{-i}^K)  +
    H(\bfY^{(i)}| \bfV_{-i}^K=\mathbf{v}_{-i}^K,\hat{\bfY}^{(i)})
  \Big]
  \\
  & \le
    nH(Y^{(i)}| V_{-i}^K)  + n \nextep +
  \sum_{\mathbf{v}_{-i}^K}
  \Big[
    \pr(\bfV_{-i}^K=\mathbf{v}_{-i}^K)
    H(\bfY^{(i)}| \bfV_{-i}^K=\mathbf{v}_{-i}^K,\hat{\bfY}^{(i)})
  \Big]
  \label{eqn:kic-equivocation-t-1}
\end{align}
Combining Fano's inequality and the fact that
\begin{equation}
  \pr(\bfY^{(i)} \neq \hat{\bfY}^{(i)}) \le \pr\Big\{ (\mathbf{V}_{-i}^K,\mathbf{Y}^{(i)})
  \not\in T_{\epsilon}^{(n)} (P_{V_{-i}^K,Y^{(i)}})\Big\}
\end{equation}
is arbitrarily small for sufficiently large $n$,
\eqn{eqn:kic-equivocation-t-1} implies
\begin{align}
  H(\bfY^{(i)}|\bfV_{-i}^K)  \le n H(Y^{(i)}|V_{-i}^K) +  n \nextepnu + n \nextep
  \label{eqn:kic-equivocation-4}
\end{align}

Substituting \eqn{eqn:kic-equivocation-1},
\eqn{eqn:kic-equivocation-2},
\eqn{eqn:kic-equivocation-3}, and \eqn{eqn:kic-equivocation-4}  into
\eqn{eqn:kic-equivocation}, we conclude that
\begin{align}
  H(W_i|\bfY^{(i)},\bfvmi^K)  \ge H(W_i) - n \epsilon^{(i)}
\end{align}
where $\epsilon^{(i)}$ is small for sufficiently large $n$,
which completes the proof.

\vspace*{-0.15in}


\end{document}